\documentclass[
,draft            % use draft while you are working on the paper
  ,numberedheadings % uncomment this option for numbered sections
  ]
  {aipproc}

\layoutstyle{6x9}
    
\begin{document}

\title{Diffraction with CDF II at the Tevatron}

%\classification{<Replace this text with PACS numbers; choose from this list: \texttt{http://www.aip.org/pacs/index.html}>}

\author{Konstantin Goulianos\footnote{Presented for the CDF II Collaboration at DIFFRACTION 2008, Sep. 9-14, La-Londe-des-Maures, France; to be published by the American Institute of Physics.}}{address={The Rockefeller University, 1230 York Avenue, New York, NY 10065, USA}}

%\author{<author2>}{address={<common address for author2 and author3>}}

%\author{<author3>}{address={<common address for author2 and author3>},altaddress={<author1 address>} % additional visiting address}

\begin{abstract}
Results on diffraction from the Fermilab Tevatron collider obtained by the CDF~II Collaboration using data from $p\bar p$ collisions at $\sqrt s=$1.96~TeV are reviewed and compared with theoretical expectations. Implications for predictions of exclusive Higgs boson production rates at the Large Hadron Collider are discussed.
\end{abstract}

\maketitle

%%%%%%%%%%%%%%%%%%%%%%%%%%%%%%%%%%%%%%%%%%%%
%% MAINMATTER
%%%%%%%%%%%%%%%%%%%%%%%%%%%%%%%%%%%%%%%%%%%%

\section{Introduction}

\begin{figure}[h]
\centerline{\includegraphics[width=0.7\textwidth]{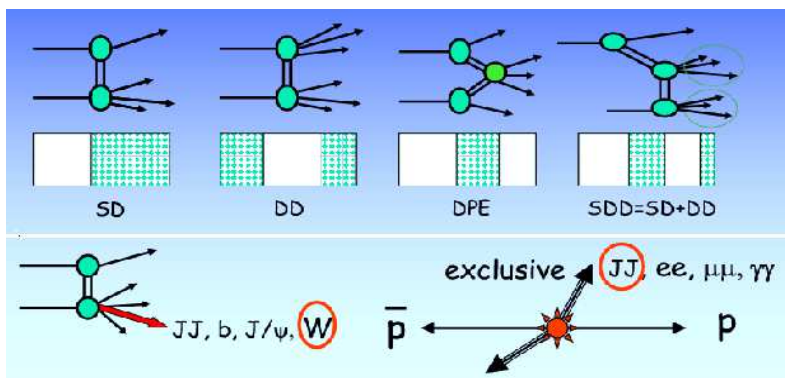}}
\caption{Diffraction at CDF: schematic diagrams and $\phi$ vs. $\eta$ topologies for 
SD$\equiv$single diffraction, DD$\equiv$double diffraction, DPE$\equiv$double Pomeron exchange, SDD$\equiv$SD$\cdot$DPE, and exclusive final state production. The hatched areas in the $\phi$-$\eta$ plots represent regions where particle production occurs.}
\label{Fig:diffraction}
\end{figure}
The phenomenal success of the standard model is tainted by our inability to analytically calculate ``soft'' processes, where the strong coupling constant is large and perturbative techniques fail. Yet, a fundamental understanding of soft processes may help reveal the underlying mechanism of confinement and explain the intricacies of hadron structure. Experimentally, diffractive processes may be used as tools for new discoveries, as they provide low background environments for certain production channels, among which most notable is the exclusive production of Higgs bosons (see Sec.~\ref{sec:dijets}).

The CDF collaboration has reported several results on soft and hard diffraction processes from $\bar pp$ collisions at the Fermilab Tevatron using rapidity gaps and~/~or a leading antiproton as a signature for diffraction (Fig.~\ref{Fig:diffraction}). These results have revealed regularities in the data that point to a QCD picture of diffraction as an exchange of a spin zero color-singlet combination of gluons and~/~or quarks carrying the quantum numbers of the vacuum (see review in \cite{lathuile07}). 

One result that has attracted widespread attention is the observation of a breakdown of QCD factorization in hard diffractive processes, which is expressed as a suppression by a factor of $\cal{O}$(10) of the production cross section relative to theoretical expectations. However, of equal importance is the finding of a breakdown of Regge factorization in soft diffraction by a factor of the same magnitude~\cite{lathuile07}. Combined, these two results strongly support the hypothesis that the breakdown of factorization is due to a saturation of the probability of forming a rapidity gap by an exchange of a color-neutral construct of the underlying parton distribution function (PDF) of the proton, which is historically referred to as ``Pomeron''. Renormalizing the ``gap probability'' to unity over all $(\xi,t)$ phase space corrects for the unphysical effect of overlapping diffractive rapidity gaps and leads to an agreement between theory and experiment (see~\cite{lathuile07} and references therein). 

The gap probability renormalization model is further supported by the following soft-diffraction results obtained by CDF~\cite{lathuile07}:
\begin{itemize}
\vspace*{-0.5em}
%\addtolength{\itemsep}{-0.75em}
\item double-diffraction (central gap): same suppression factor as in single-diffraction;  
\item {multi-gap diffraction:} double-gap to single-gap ratio non-suppressed;
\item {energy independence:} $\sigma^D_{tot}\rightarrow$~constant as $s\rightarrow\infty$;
\item {Pomeron intercept and slope:} they are related!~\cite{blois07}.
\end{itemize}
Similar results are found for hard-diffraction. 

In this paper, we review the results obtained in Run~II.
\begin{figure}[h]
\vspace*{-1em}
\centerline{\includegraphics[width=0.8\textwidth]{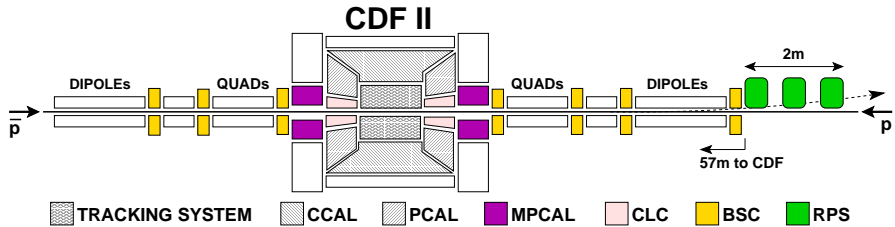}}
\caption{The CDF II detector.}\label{Fig:detector}
\end{figure}
The CDF~II detector is shown schematically in Fig.~\ref{Fig:detector} (from Ref.~\cite{excl2j}). The components of the main detector~\cite{cdf2detector} used in the diffractive program are the tracking system, the central (CCAL), plug (PCAL), and forward (FCAL) calorimeters, and the \v{C}erenkov luminosity counters (CLC). 

The diffractive program benefited from dedicated triggers and a system of special forward detectors.
The following forward detectors were employed~\cite{excl2j}:   
\begin{itemize}
%\addtolength{\itemsep}{-0.5em}
\item  RPS (Roman Pot Spectrometer)~-~detects leading $\bar p$'s at $\sim 0.03<\xi\equiv 1-p_{||}<0.09$;
\item MPCAL (MiniPlug Calorimeters)~-~measure $E_T$ and $(\theta,\phi)$ at $\sim 3.5<|\eta|<5.5$;
\item BSC (Beam Shower Counters)~-~identify rapidity gaps at $\sim 5.5<|\eta|<7.5$.
\end{itemize}
There are three classes of results obtained thus far from Run~II data:
\begin{itemize}
\item Exclusive dilepton and diphoton production (see talk by M. Albrow).
\item Rapidity gaps between jets (see talk by C. Mesropian).
\item Diffractive / Exclusive dijet and W / Z production (this talk).
\end{itemize}

\section{Diffractive W / Z production}
Diffractive dijet production at the Tevatron is suppressed by a factor of $\cal{O}$(10) relative to  expectations based on the proton PDF extracted from diffractive deep inelastic scattering (DDIS) at the DESY $ep$ Collider HERA (see Ref.~\cite{lathuile07}. While  
 no DDIS suppression is expected in certain models (see e.g.~\cite{collins}), the primary exchange in DDIS is a $q\bar q$ pair, while dijets are mainly produced by a $gg$ exchange. The dijet rates at the Tevaytron are calculated using a gluon PDF extracted from DDIS. A more direct comparison could be made by measuring the DSF in diffractive $W$ production at the Tevatron, which is dominated by a $q\bar q$ exchange as in DDIS. In Run~I, only the overall diffractive $W$ fraction was measured~\cite{cdf_W}. In Run~II, CDF measured both the $W$ and $Z$ diffractive fractions and also lpans to attempt to measure the DSF.

\begin{figure}[h]
\centerline{\includegraphics[width=1.0\textwidth]{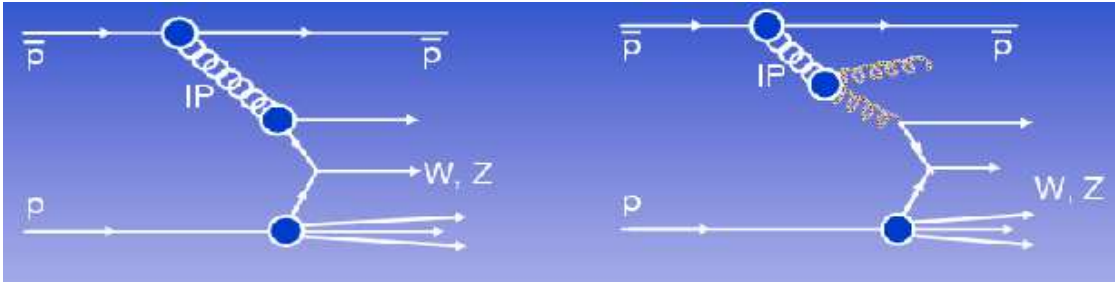}}
\caption{LO diffractive $W/Z$ production diagrams.}\label{wz_diagrams}
\vspace*{-1em}
\end{figure}
\noindent Figure~\ref{wz_diagrams} shows schematic Fynman diagrams for diffractive $W/Z$ production. In leading order, the $W/Z$ is produced by a quark in the Pomeron (left), while production by a gluon (right) is suppressed by a factor of $\alpha_s$ and can be distinguished from quark production by an associated jet~\cite{cdf_W}. 

The data analysis is based on events with RPS tracking from a data sample of approximately $0.6$~fb$^{-1}$. In addition to the $W/Z$ selection requirements (see below), a hit in the RPS trigger counters and a RPS reconstructed track with $0.03<\xi<0.1$ and $|t|<1$ are required. 
%Details can be found in the CDF public web page~\cite{cdfweb_W}. 
A novel feature of the analysis is the determination of the full kinematics of the $W\rightarrow e\nu/\mu\nu$ decay, which is made possible  by obtaining the neutrino $E_T^\nu$ from the missing $E_T$, as usual, and $\eta_\nu$ from the formula $\xi^{\rm RPS}-\xi^{\rm cal}=(E_T/\sqrt{s})\exp[-\eta_\nu]$ , where $\xi^{\rm cal}=\sum_{\rm towers}(E_T/\sqrt{s})\exp[-\eta]$. 

\begin{figure}[ht]
\includegraphics[width=0.5\textwidth]{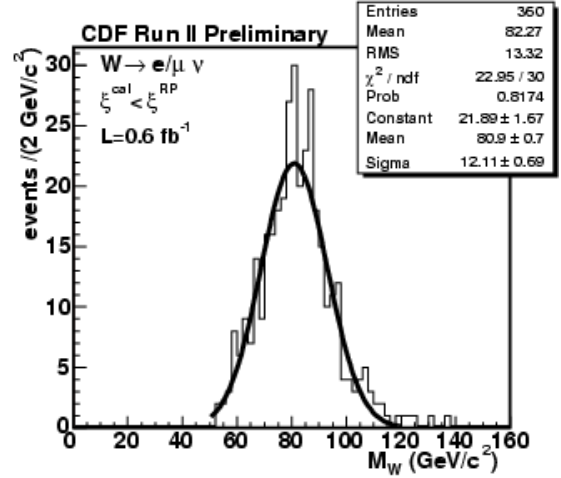}
%\vspace*{-2em}
\caption{Histogram of the $W$ mass from the diffractive data sample and a Gaussian fit.}
\label{Fig:W-mass}
\end{figure}

The CDF $W/Z$ selection requirements are $E_T^{e,\mu}>25$~GeV, $40<M_T^W<120$~GeV, $66<M^Z<116$~GeV, and vertex $z$-coordinate $z_{vtx}<60$~cm. The $W$ mass distribution for events with $\xi^{\rm CAL}<\xi^{RPS}$ is shown in Fig.~\ref{Fig:W-mass} along with a Gaussian fit. The obtained value of $M_W^{\rm exp}=80.9\pm 0.7$~GeV is in good agreement with the world average $W$-mass of $M_W^{\rm PDG}=80.403\pm 0.029$~GeV~\cite{PDG}. 

\begin{figure}[h]
\centerline{\includegraphics[width=0.4\textwidth]{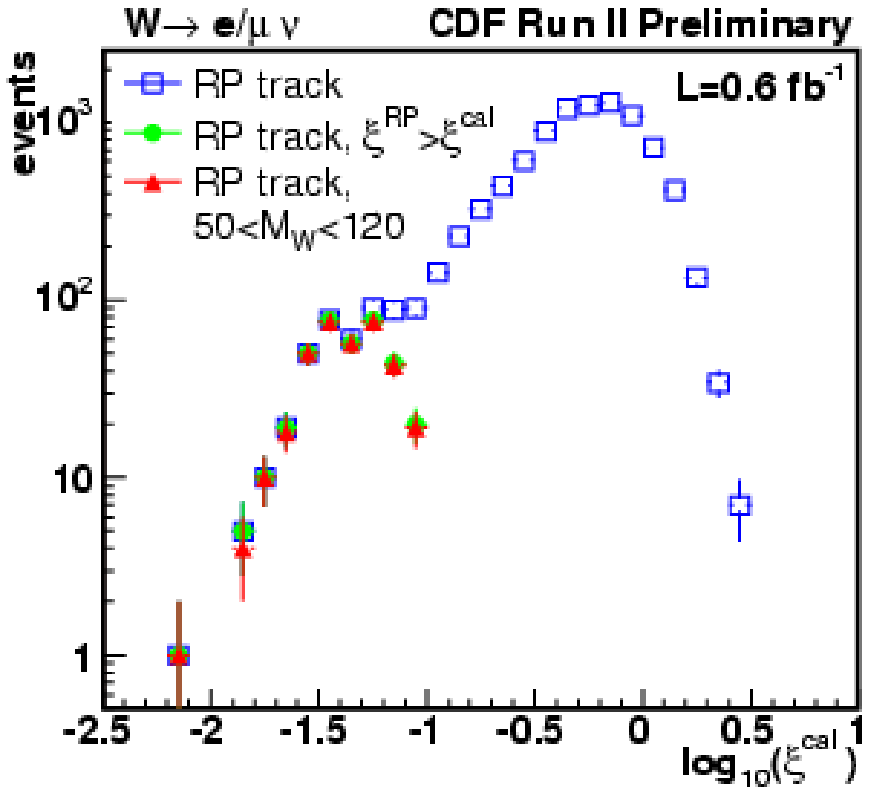}
\includegraphics[width=0.415\textwidth]{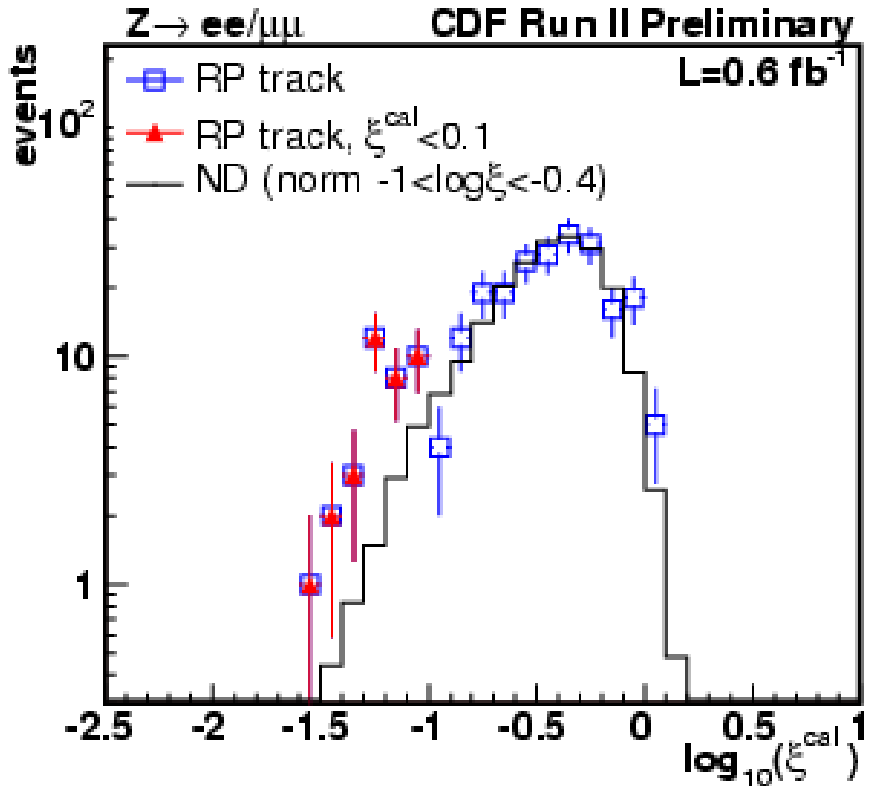}}
\vspace*{-1em}
\caption{The $\xi^{\rm CAL}$ distribution for various $W$ (left) and $Z$ (right) event samples.}
\label{Fig:xi_cal}
\end{figure}

Figure~\ref{Fig:xi_cal} shows the $\xi^{\rm CAL}$ distributions of the $W/Z$ events satisfying different selection requirements. In the $W$ case, the requirement of $\xi^{\rm RP}>\xi^{\rm CAL}$ is very effective in removing the overlap events in the region of $\xi^{\rm CAL}<0.1$, while a mass cut of $50<M_W<120$~GeV has the same effect. In the $Z$ case, the $\xi^{\rm CAL}$ distribution of all $Z$ events is used and normalized to the RP-track distribution in the region of $-1<\log\xi^{\rm CAL}<-0.4$ ($0.1<\xi^{\rm CAL}<0.4$) to obtain the ND background in the diffractive region of $\xi^{\rm CAL}<0.1$.   
 
Accounting for the RPS acceptance of $A_{\rm RPS}\approx 80$~\%, the trigger counter efficiency of $\epsilon_{\rm RPStrig}\approx 75$~\%, the track reconstruction efficiency of $\epsilon_{\rm RPStrk}\approx 87$~\%, multiplying by two to include the production by $\bar pp\rightarrow X+W/Z+p$, and correcting the number of ND events for the effect of overlaps due to multiple interactions by multiplying it by a factor of $f_{\rm 1-int}\approx 25$~\%, allows the calculation of the diffractive fraction of $W/Z$ events as 

$R_{W/Z}=2\cdot N_{SD}/A_{\rm RPS}/\epsilon_{\rm RPStrig}/\epsilon_{\rm RPStrk}/(N_{\rm ND}\cdot f_{\rm 1-int})$, which yields the results:  
\vglue 1em      
$R_W(0.03<\xi<0.10,\,|t|<0.1)=[0.97\pm 0.05\;\mbox{(stat)}\pm 0.11\;\mbox{(syst)]}\%,$

$R_Z(0.03<\xi<0.10,\,|t|<0.1)=[0.85\pm 0.20\;\mbox{(stat)}\pm 0.11\;\mbox{(syst)]}\%.$
\vglue 1em

\noindent The $R_W$ value is consistent with the Run~I result of:

$R_W(0.03<\xi<0.10,\,|t|<0.1)=[0.97\pm 0.47]~\%\;(Run~I)$, 

\noindent obtained from the measured value of $R^W(\xi<0.1)=[0.15\pm0.51\;\mbox{(stat)}\pm0.20\;\mbox{(syst)}]\%$~\cite{cdf_W}, which is multiplied by a factor of 0.85 tto account for the reduced ($\xi$-$t$) range in Run~II.
\section{Diffractive and Exclusive dijet production}\label{sec:dijets} 
\begin{center}
\begin{figure}[h]
\hspace*{2em}\includegraphics[width=0.51\textwidth]{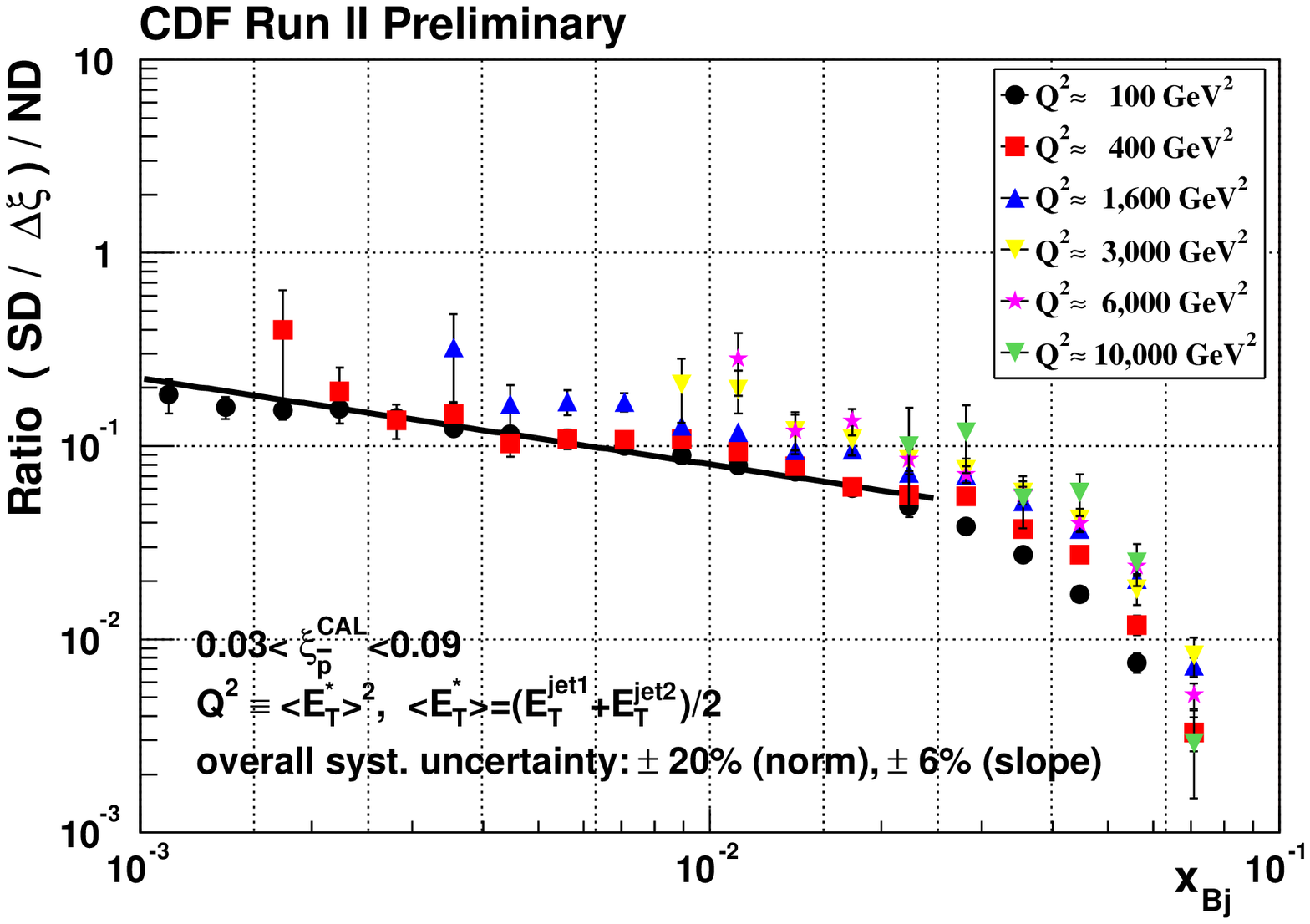}

\includegraphics[width=0.49\textwidth]{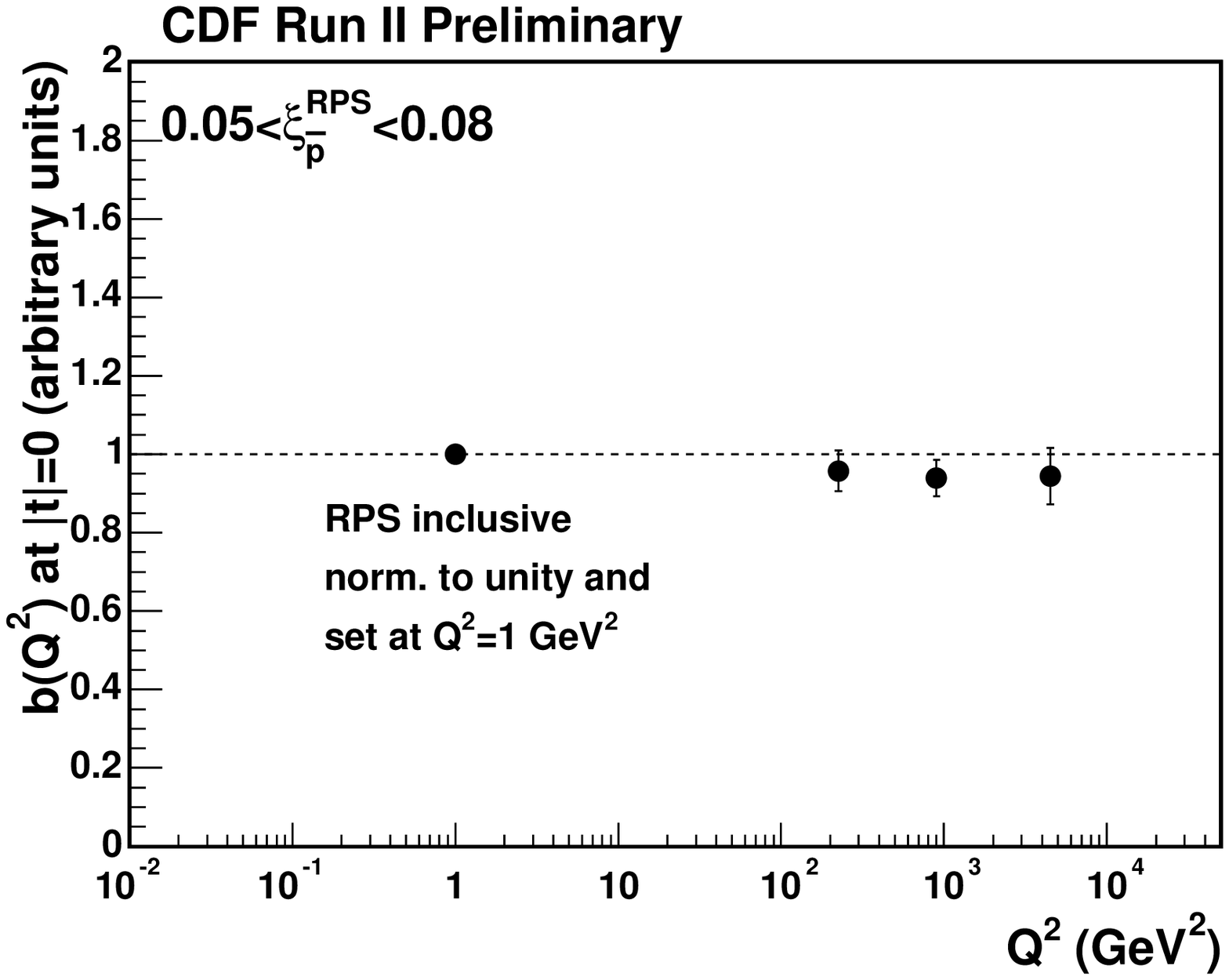}
\caption{
{\em(left)} Ratio of single-diffractive (SD) to non-diffractive (ND) dijet event rates vs. $x_{Bj}$ (momentum fraction of parton in $\bar p$) for different value
s of $E_T^2\equiv Q^2$;
{\em (right)} $b|_{t=0}$ slope vs. $Q^2$.
}
\label{fig:xbjQ2}
\end{figure}
\end{center}

\paragraph{Diffractive dijet production}
Preliminary resluts on the $x_{Bj}$, $Q^2$, and $t$ dependence of the diffractive structure function from dijet production at $\sqrt s=1960$ GeV were reported at ``Diffraction-2006''~\cite{PoS2006cdf}~\footnote{This section is an excerpt from Ref.~\cite{PoS2006cdf}.}. The measured Run~II $x_{Bj}$ rates confirm the factorization breakdown observed in Run~I. The $Q^2$ and $t$ dependence results are shown in Fig~\ref{fig:xbjQ2}.
\vspace*{-1em}
{\bf $Q^2$-dependence.} In the range  $10^2\hbox{ GeV}^2<Q^2<10^4$~GeV$^2$, where the inclusive $E_T$ distribution falls by a factor of $\sim 10^4$, the ratio of the SD/ND distribution increases by only a factor of $\sim 2$. The above results indicate that the $Q^2$ evolution in diffractive interactions is similar to that in ND interactions.

\noindent {\bf $t$-dependence.} The slope parameter $b(Q^2)|_{t=0}$ of an exponential fit to $t$ distributions near $t=0$ shows no $Q^2$ dependence in the range $1\hbox{ GeV}^2<Q^2<10^4\hbox{ GeV}^2$.

 The above results support the picture of a composite Pomeron formed from color-singlet combinations of the underlying parton densities of the nucleon~\cite{lathuile07}.

\begin{figure}[ht]
 \includegraphics[width=3cm]{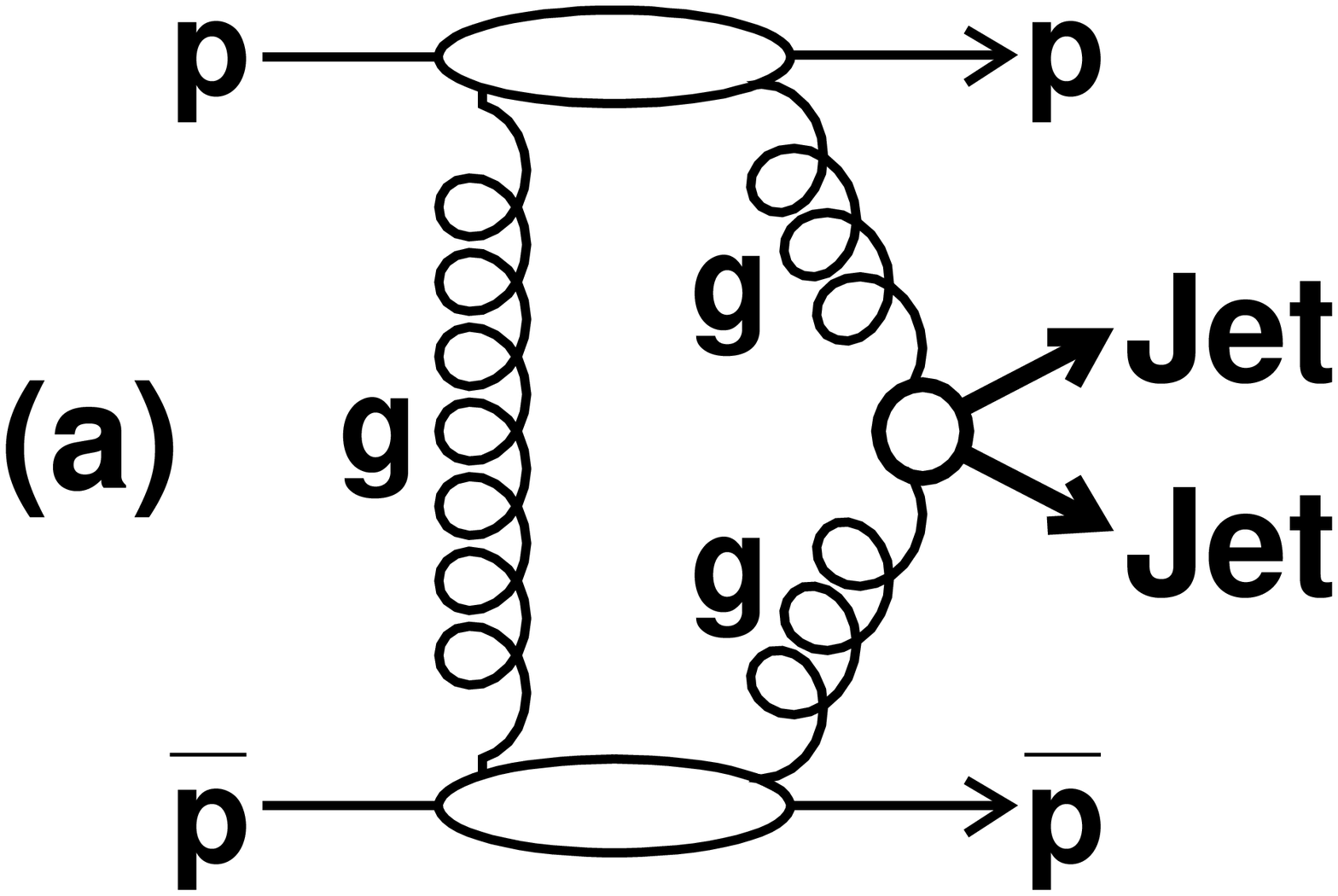}
 
\includegraphics[width=3cm]{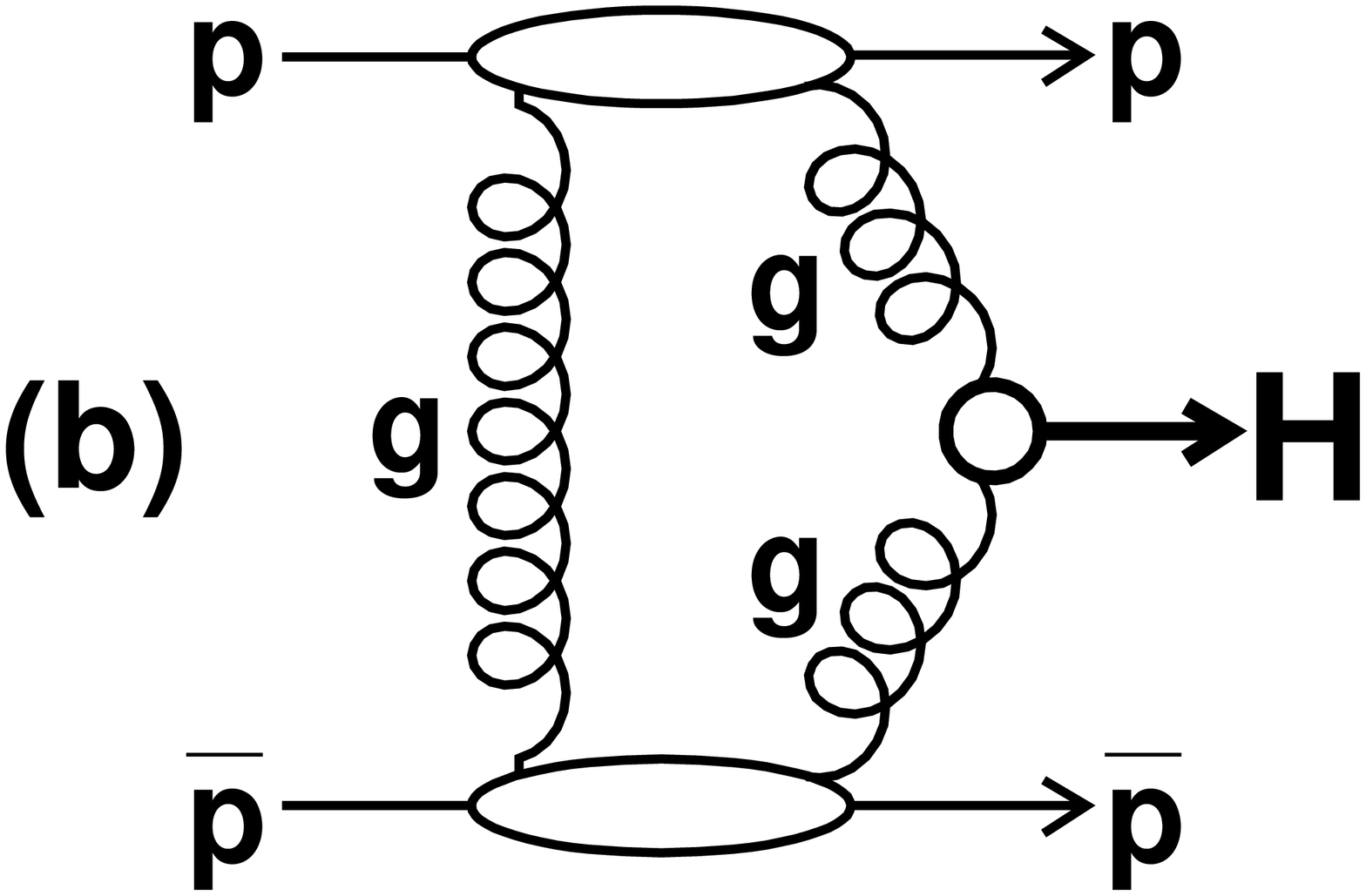}

\caption{Diagrams for (a) exclusive dijet and (b) Higgs boson production.}
\label{Fig:xsec_vs_mjj}
\end{figure}

\paragraph{Exclusive dijet production}
The process of exclusive dijet production is important for testing and~/~or calibrating models for exclusive Higgs production at the LHC. The CDF collaboration has made the first observation of this process and their main result is shown in Fig.~\ref{Fig:xsec_vs_mjj}. Details can be found in Ref.~\cite{excl2j}. This result favors the model of Ref.~\cite{KMR}, which is implemented in the Monte Carlo simulation {\sc ExHuME}~\cite{ExHuME}.
\begin{figure}[ht]
\includegraphics[width=8cm]{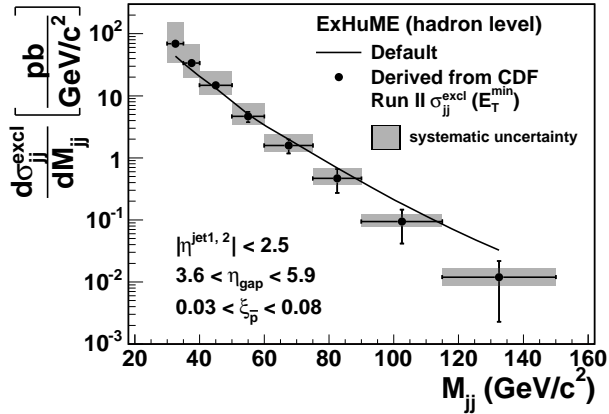}

\caption{{\sc ExHuME}~\cite{ExHuME} exclusive dijet differential cross sections at the hadron level vs. dijet mass $M_{jj}$ normalized to measured $\sigma_{jj}^{excl}$ values. The solid curve is the cross section predicted by {\sc ExHuME}.
%Exclusive dijet production cross sections vs. dijet mass $M_{jj}$ derived from events generated by the {\sc ExHuME} Monte Carlo~\cite{ExHuME} simulation using a procedure that calibrates the MC using CDF data~\cite{excl2j}. The solid curve is the cross section predicted by {\sc ExHuME} using the default settings.
}
\end{figure}

\section{Conclusion}
We review CDF~II results from $p\bar p$ collisions at $\sqrt s=$1.96~TeV and compare them with theoretical expectations. We concentrate on diffractive W~/~Z and dijet production, a comparison of which allows the determination of the quark to gluon ratio of the diffractive exchange. We also discuss the recently published result of exclusive dijet production, which is used to check~/~calibrate theoretical predictions for exclusive Higgs boson production at the LHC. Results from CDF~II on rapidity gaps between jets and on exclusive diphoton and dilepton production have also been presented at this conference.
Combined, these results and the analysis methods used in obtaining them provide a powerful launching board for searches at the LHC aimed at exploiting diffractive and exclusive production to discover new physics.

\begin{footnotesize}
% IF YOU DO NOT USE BIBTEX, USE THE FOLLOWING SAMPLE SCHEME FOR THE REFERENCES
% ----------------------------------------------------------------------------

% -------------------------------------------------------------------
\end{footnotesize}

% ****************************************************************************
% END OF BIBLIOGRAPHY AREA
% ****************************************************************************

\end{document}